\documentclass[aps,reprint,showpacs,floatfix]{revtex4-1}
\usepackage{amsmath,graphics,epsfig,color,verbatim,gensymb,epstopdf}
\let\oldhat\hat
\renewcommand{\vec}[1]{\mathbf{#1}}
\renewcommand{\hat}[1]{\oldhat{\mathbf{#1}}}

\begin{document}

%\title{Density functional plus dynamical mean-field theory of Li$_x$CoO$_2$}
\title{Compositional phase stability of correlated electron materials within DFT+DMFT}
\author{Eric B. Isaacs}
\altaffiliation[Present address: ]{Department of Materials Science and Engineering, Northwestern University, Evanston, IL 60208}
%% \email{eric.isaacs@northwestern.edu}
\affiliation{Department of Applied Physics and Applied Mathematics, Columbia University, New York, NY 10027}
\author{Chris A. Marianetti}
\email{chris.marianetti@columbia.edu}
\affiliation{Department of Applied Physics and Applied Mathematics, Columbia University, New York, NY 10027}

\begin{abstract}
Predicting the compositional phase stability of strongly correlated
electron materials is an outstanding challenge in condensed matter
physics, requiring precise computations of total energies. In this
work, we employ the density functional theory plus dynamical
mean-field theory (DFT+DMFT) formalism to address local correlations
due to transition metal $d$ electrons on compositional phase stability
in the prototype rechargeable battery cathode material Li$_x$CoO$_2$,
and detailed comparisons are made with the simpler DFT+$U$ approach
(i.e., the Hartree-Fock solution of the DMFT impurity problem). Local
interactions are found to strongly impact the energetics of the band
insulator LiCoO$_2$, most significantly via the $E_g$ orbitals, which
are partially occupied via hybridization with O $p$ states. We find
CoO$_2$ and Li$_{1/2}$CoO$_2$ to be moderately correlated Fermi
liquids with quasiparticle weights of 0.6--0.8 for the $T_{2g}$
states, which are most impacted by the interactions. As compared to
DFT+$U$, DFT+DMFT considerably dampens the increase in total energy as
$U$ is increased, which indicates that dynamical correlations are
important to describe this class of materials despite the relatively
modest quasiparticle weights. Unlike DFT+$U$, which can incorrectly
drive Li$_x$CoO$_2$ towards spurious phase separating or charge
ordered states, DFT+DMFT correctly captures the system's phase
stability and does not exhibit a strong charge ordering tendency. Most
importantly, the error within DFT+$U$ varies strongly as the
composition changes, challenging the common practice of artificially
tuning $U$ within DFT+$U$ to compensate the errors of Hartree-Fock.
DFT+DMFT predicts the average intercalation voltage decreases relative
to DFT, \emph{opposite} to the result of DFT+$U$, which would yield
favorable agreement with experiment in conjunction with the
overprediction of the voltage by the strongly constrained and
appropriately normed (SCAN) DFT functional.

%
%With respect to the
%DFT prediction, the change in the average intercalation voltage
%predicted by DFT+DMFT %based on the generalized gradient approximation of DFT
%is \emph{opposite} to that of DFT+$U$, decreasing the
%voltage relative to DFT
%

 %is complementary to  %Given that the more advanced strongly constrained

%overpredicts the
%voltage, a decrease in voltage predicted
%by DFT+DMFT will lead to improved agreement between theory and
%experiment within DFT(SCAN)+DMFT.

%% which is commensurate with the fact that the
%% more advanced SCAN DFT functional overpredicts the voltage.

%% XXX We may want to remove mention of $N_d$ altogether so as to not have to mention charge self-consistency in abstract.

%% In both systems, there are
%% appreciable dynamical charge and spin fluctuations.

%% and leads to more modest
%% changes in the number of $d$ electrons ($N_d$) as a function of $U$
%% (when excluding full charge self-consistency).

%% and when possible compare results for projector
%% versus maximally-localized Wannier function correlated subspace
\end{abstract}

\date{\today}
%% \pacs{71.15.Mb, 71.27.+a, 81.30.Bx, 82.47.Aa}
\maketitle

\section{Introduction}

Strongly correlated materials, for which density functional theory
(DFT) calculations often break down due to strong electron-electron
interactions, are a challenging class of condensed matter systems
relevant to several important technologies
\cite{kotliar_strongly_2004,morosan_strongly_2012}. One example is
Li ion rechargeable batteries. These electrochemical cells rely
critically on a cathode material that can reversibly intercalate Li
ions \cite{whittingham_lithium_2004}. Since cathode materials
typically are based on transition metal oxides to accommodate changes
in oxidation state, they have an open $d$ electron shell and are
susceptible to strong correlation physics.

Currently, the dominant cathode materials are based on Li$_x$CoO$_2$
(LCO), a layered compound in which Li ions are intercalated between
layers of edge-sharing Co--O octahedra, as shown in Fig.
\ref{fig:dos_structure}(d) \cite{mizushima_lixcoo2_1980}. Several
early theoretical studies that revealed significant insight into the
electronic structure and phase diagram of LCO
\cite{czyzyk_band-theory_1992,aydinol_ab_1997,wolverton_first-principles_1998,van_der_ven_first-principles_1998}
were based on DFT
\cite{hohenberg_inhomogeneous_1964,kohn_self-consistent_1965}, the
\emph{de facto} standard for first-principles calculations in
solid-state physics and chemistry. It is not uncommon, however, for
DFT to fail to capture the physics of correlated materials due to the
approximation for the exchange-correlation functional [e.g., local
  density approximation (LDA) or generalized gradient approximation
  (GGA)].

%(1) the approximation of the effective
%potential [e.g. local density approximation (LDA) or generalized
%  gradient approximation (GGA)] and (2) description in terms of
%single-particle Kohn-Sham states.

While DFT calculations in many ways reliably characterizes LCO, there are
deficiencies in their description. DFT (GGA) underestimates the intercalation
voltage by around 0.8 V \cite{chevrier_hybrid_2010}. In addition,
using a plane-wave basis set and ultrasoft pseudopotentials, Van der
Ven \textit{et al.} found that LDA overestimates the order-disorder
transition temperature for $x=1/2$ by 100 \degree C
\cite{van_der_ven_first-principles_1998}. We note that the LDA linear
augmented plane wave results of Wolverton and Zunger
\cite{wolverton_first-principles_1998} do not show the same
overestimation, though this study uses only roughly one third as many DFT calculations to parameterize the cluster expansion, in addition
to performing slightly restricted structural relaxations.

One widely utilized approach to go beyond DFT is the DFT+$U$ method
\cite{liechtenstein_density-functional_1995}, in which an explicit
on-site Coulomb interaction $U$ is added to account for the strong
interactions in the $d$ shell along with a simple mean-field ansatz
for the energy functional. However, DFT+$U$ does not fully remedy the
shortcomings of DFT and in some cases hurts the description more than
it helps. DFT+$U$ still underestimates the voltage by 0.3 V, and it
can overestimate the order-disorder transition temperature by as much
as several hundred degrees \cite{isaacs_compositional_2017}. DFT+$U$
drives LiCoO$_2$ towards a high-spin transition
\cite{andriyevsky_electronic_2014} not observed in experiments
\cite{van_elp_electronic_1991,menetrier_insulator-metal_1999,menetrier_really_2008}
and, unless spurious charge ordering is permitted to occur,
incorrectly predicts phase separation
\cite{isaacs_compositional_2017,seo_calibrating_2015}. Moreover,
DFT+$U$ finds CoO$_2$ to be an insulator in disagreement with
experiment \cite{zhang_doping_2004}. DFT+$U$ clearly is problematic in
the context of LCO.

Here, we revisit the electronic structure, voltage, and phase
stability of LCO using more sophisticated DFT plus dynamical
mean-field theory (DFT+DMFT) calculations
\cite{kotliar_electronic_2006} based on GGA. In this framework, the
many-body DMFT approach captures the dynamical local correlations of
Co $d$ electrons embedded in the crystal, whereas only the static
effects are described within DFT+$U$. Total energy DFT+DMFT
calculations have become an important tool for understanding
structural stability of materials with electronic correlations
\cite{savrasov_correlated_2001,held_cerium_2001,dai_calculated_2003,mcmahan_combined_2005,amadon_alpha_2006,pourovskii_self-consistency_2007,leonov_structural_2008,di_marco_correlation_2009,leonov_computation_2010,kunes_dynamical_2009,aichhorn_importance_2011,leonov_calculated_2012,grieger_approaching_2012,bieder_thermodynamics_2014,zaki_failure_2014,park_computing_2014,park_total_2014,chen_density_2015,haule_free_2015,delange_large_2016,haule_mott_2017,haule_structural_2018,han_phonon_2018},
and our work extends this exploration to the realm of compositional
phase stability.

We find that DFT+DMFT describes LiCoO$_2$ as a band insulator with
modest shifts and broadenings of the low-energy spectrum, most
prominently via the $E_g$ levels partially occupied via hybridization
with O $p$ states. CoO$_2$ and Li$_{1/2}$CoO$_2$ are Fermi liquids
whose $T_{2g}$ states are most strongly affected by the interactions,
with quasiparticle weight of around 0.6--0.7. DFT+DMFT, unlike
DFT+$U$, does not strongly stabilize charge ordering in
Li$_{1/2}$CoO$_2$, nor does it predict insulating behavior for CoO$_2$
or Li$_{1/2}$CoO$_2$; in other words, DFT+DMFT substantially improves
the description of the electronic structure. Dynamical correlations
significantly dampen the impact of $U$ on the total energy of LCO, but
more substantially for CoO$_2$ than LiCoO$_2$, leading to a reduction
in voltage as compared to DFT, whereas DFT+$U$ yields the
qualitatively opposite behavior. Given the more accurate strongly
constrained and appropriately normed (SCAN) DFT functional
overestimates the experimental voltage, such a decrease in the
predicted voltage is expected to lead to agreement between
experimental and predicted voltage for DFT+DMFT based on the SCAN
functional. Similar to the voltage behavior, the $x=1/2$ formation
energy prediction is significantly affected by dynamical correlations:
unlike DFT+$U$, DFT+DMFT only weakly influences the formation energy
of $x=1/2$ as compared to DFT. Our results demonstrate the importance
of dynamical correlations, missing in DFT+$U$, to accurately describe
the electronic structure and energetics of correlated electron
materials.

%% Not sure exactly what to say about formation energy since the sign
%% of DFT+U correction (w.r.t. DFT) is different for Wannier and
%% projector.

%% These significant differences in the predicted voltage and
%% formation energy indicate that dynamical correlations, missing in
%% DFT+$U$, will be necessary for accurate total energy predictions in
%% correlated electron materials.

%% Finally, we
%% discuss the impact of charge self-consistency, the effect of different
%% choices of the correlated subspace, and possible reasons for the
%% underestimation of the experimental voltage within our DFT+DMFT
%% results.

\section{Computational Details}

We perform single-site paramagnetic DFT+DMFT total energy calculations
using the formalism of Ref. \citenum{park_computing_2014} based on the
spin-independent GGA of Perdew, Burke, and Ernzerhof (PBE)
\cite{perdew_generalized_1996} and the projector augmented wave (PAW)
method \cite{blochl_projector_1994,kresse_ultrasoft_1999} as
implemented in the \textsc{vasp} code
\cite{kresse_ab_1994,kresse_ab_1993,kresse_efficient_1996,kresse_efficiency_1996}.
Select calculations are also performed using the SCAN functional
\cite{sun_strongly_2015}. Given that LCO exhibits no long range
magnetic order in experiment \cite{motohashi_electronic_2009}, the
paramagnetic state is justified and we do not search for long-range
magnetic order. The structures are fixed to the fully relaxed
spin-dependent DFT ground state structures with O3 layer stacking
\cite{van_der_ven_first-principles_1998}, corresponding to a band
insulator for $x=1$ and ferromagnetic low-spin metals for $x=0$ and
$x=1/2$. Except where otherwise noted, calculations are performed
using the fixed non-spin-polarized DFT charge density, i.e., they are
non-charge-self-consistent (NCSC); this is done for reasons of
computational efficiency. We characterize the magnitude of the error
associated with charge self-consistency by directly comparing NCSC and
charge-self-consistent (CSC) calculations within DFT+$U$,
demonstrating that the error is sufficiently small for the trends we
are studying in this paper. A 500 eV energy cutoff and $k$-point
meshes of $k$-point density corresponding to $9\times9\times9$ for the
rhombohedral LiCoO$_2$ primitive cell and $19\times19\times19$ for the
bulk Li primitive cell are employed. The ionic forces and total energy
are converged to 0.01 eV/\AA\ and 10$^{-6}$ eV, respectively.

To define the correlated subspace, we utilize the maximally-localized
Wannier function (MLWF) basis \cite{mostofi_wannier90:_2008} for the
full $p$-$d$ manifold and perform a unitary rotation of the $d$
orbitals to minimize the off-diagonal hoppings
\cite{park_computing_2014}. The Slater-Kanamori (SK) interaction with
$J_{\mathrm{SK}}$ set to 0.7 eV is employed, and we use the
numerically exact hybridization expansion continuous-time quantum
Monte Carlo (CTQMC) solver for the 5-orbital impurity problem
\cite{haule_quantum_2007,gull_continuous-time_2011} at temperature
$T=290$ K. For DMFT, we perform calculations using density-density
interactions. In order to assess the impact of terms beyond
density-density interactions, we also perform calculations augmenting
density-density interactions with the off-diagonal $J$ terms within
the $E_g$ manifold, in which the impact of off-diagonal $J$ is
expected to be most important due to the partial filling of $E_g$. For
comparison, we also perform DFT+$U$ calculations using the projector
(corresponding to projection onto spherical harmonics within the PAW
spheres \cite{bengone_implementation_2000}) correlated subspace in
\textsc{vasp} (\textsc{ldautype=4}) and present all our results in
terms of the $U$ and $J$ corresponding to this interaction model via
$U=U_{\mathrm{SK}}-8J_{\mathrm{SK}}/5$ and $J=7J_{\mathrm{SK}}/5$
\cite{pavarini_lda+dmft_2011}. Although it has limitations such as
lack of normalization \cite{haule_dynamical_2010,schuler_charge_2018},
we consider the projector correlated subspace since it is widely used.
It should be noted that $J$ is fixed in all calculations, even though
a range of different $U$ is explored. We employ the
fully-localized-limit (FLL) form of the double counting
\cite{anisimov_density-functional_1993}. The total energy is converged
to within $\sim 10$ meV/f.u. for the DMFT self-consistency condition.

\section{Results and Discussion}
\subsection{Electronic structure of CoO$_2$ and LiCoO$_2$}
\begin{figure}[htbp]
\begin{center}
\includegraphics[width=\linewidth]{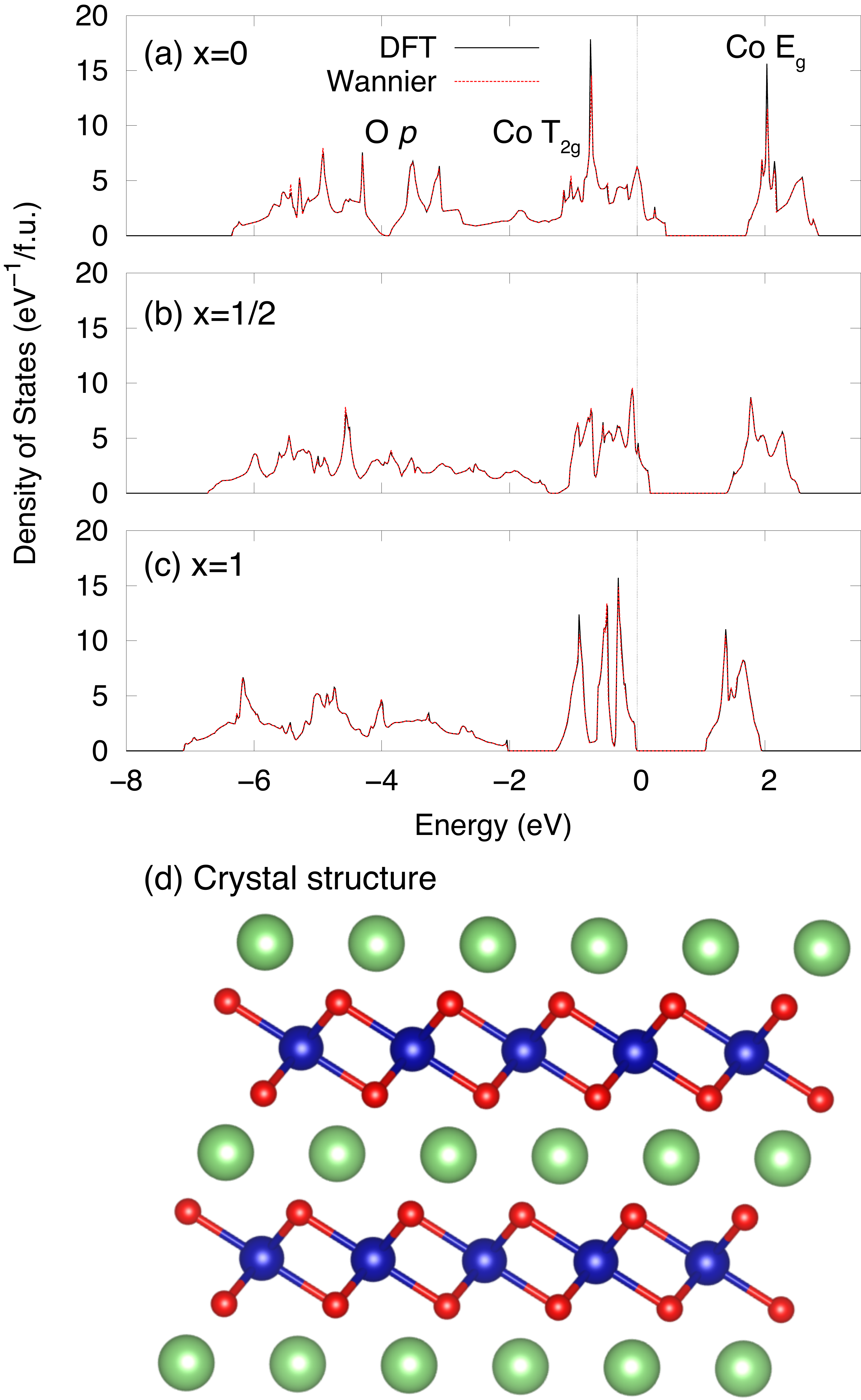}
\end{center}
\caption{Density of states for DFT (black solid lines) and using the
  Wannier basis (dotted red lines) for (a) metallic CoO$_2$, (b)
  metallic Li$_{1/2}$CoO$_2$, and (c) band insulator LiCoO$_2$. The
  Fermi level (valence band maximum for $x=1$) is indicated by the
  vertical dotted black line. (d) Crystal structure of LCO with O3
  layer stacking with all the Li shown ($x=1$). The large green,
  medium blue, and small red spheres represent ionic positions of Li,
  Co, and O, respectively. The image of the crystal structure is
  generated using \textsc{vesta}
  \cite{momma2011vesta}.\label{fig:dos_structure}}
\end{figure}

We begin by studying the basic electronic structure of LiCoO$_2$ and
CoO$_2$ as a function of $U$, allowing for a direct comparison between
DFT+DMFT and DFT+$U$. Corresponding results for Li$_{1/2}$CoO$_2$ are
included, but these are not directly discussed until Section
\ref{sec:es_x12}. In LCO, the ability of the oxygens to relax in the
out-of-plane direction slightly distorts the CoO$_6$ octahedra and
results in a symmetry lineage of $T_{2g} \rightarrow A_{1g}+E_g'$
relative to cubic symmetry, though we will still sometimes refer to
this manifold as $T_{2g}$ for brevity. The DFT density of states is
shown in Fig. \ref{fig:dos_structure}(a) and
\ref{fig:dos_structure}(c) for CoO$_2$ and LiCoO$_2$, respectively.
Within DFT, LiCoO$_2$ is a band insulator with nominally filled
$T_{2g}$ and empty $E_g$ states, whereas CoO$_2$ is metallic with a
hole in the $T_{2g}$ manifold. The density of states from the Wannier
correlated subspace for the full $p$-$d$ manifold, shown in the dashed
red lines, is numerically identical to that of DFT by construction.
The Wannier functions are well localized with values for the spread
$\langle(\vec{r}-\bar{\vec{r}})^2\rangle$ of around 0.42 and 0.45
\AA$^2$ for the individual Co $d$ orbitals of CoO$_2$ and LiCoO$_2$,
respectively.

%\subsection{Self-energy}

\begin{figure*}[htbp]
\begin{center}
\includegraphics[width=\linewidth]{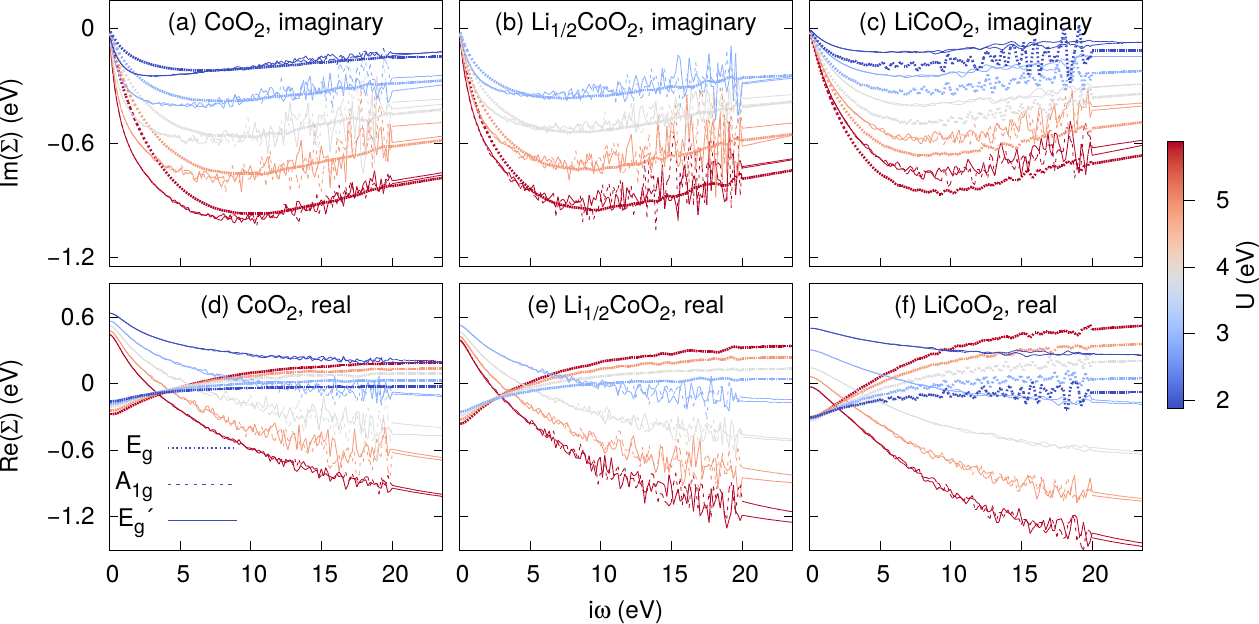}
\end{center}

\caption{Imaginary part of the DMFT self-energy on the imaginary
  frequency axis for (a) CoO$_2$, (b) Li$_{1/2}$CoO$_2$, and (c)
  LiCoO$_2$ with density-density interactions for different values of
  $U$. Solid, dashed, and dotted lines correspond to the $E_g'$,
  $A_{1g}$, and $E_g$ orbitals. (d), (e), and (f) show the
  corresponding real parts referenced to the chemical
  potential.\label{fig:sigma_densdens}}
\end{figure*}

An essential quantity in Green function based approaches is the self-energy,
which is central to computing the total energy and determining the low-energy properties.
The electronic self-energy $\Sigma$ on the imaginary (Matsubara)
frequency axis obtained via the CTQMC solver is shown for CoO$_2$ and
LiCoO$_2$ in Fig. \ref{fig:sigma_densdens} for density-density
interactions. The noise in the self-energy stems from the stochastic
nature of the CTQMC solver, and for frequencies above 20 eV there is
no noise since we utilize the analytic form of $\Sigma$ in the
high-frequency limit. We note that the self-energy is well converged,
particularly for low frequency.

For both CoO$_2$ and LiCoO$_2$, $\mathrm{Im}(\Sigma)$ goes to 0 at low
frequency, consistent with well-defined quasiparticles and a band
insulator, respectively. This indicates that CoO$_2$ can be described
as a Fermi liquid and is not a Mott insulator, consistent with
experiments on CoO$_2$
\cite{de_vaulx_electronic_2007,motohashi_synthesis_2007,kawasaki_measurement_2009},
whereas past DFT+$U$ studies predict an insulating state
\cite{zhang_doping_2004}; as do our DFT+$U$ results in the present study. Therefore, DFT+DMFT is providing an improved
description of the electronic structure of LCO. As a function of $U$,
the magnitude of $\mathrm{Im}(\Sigma)$ increases. The imaginary part
of the self-energy is essentially identical for the $E_g'$ and
$A_{1g}$ states, which indicates the symmetry breaking within the
$T_{2g}$ manifold is small. The overall magnitude of
$\mathrm{Im}(\Sigma)$ is moderately larger for CoO$_2$ than for
LiCoO$_2$. For CoO$_2$, the imaginary part of the self-energy of the
$E_g'$ and $A_{1g}$ states are larger in magnitude than those of the
$E_g$ states below $i\omega\approx 10$ eV, which is reasonable given
their respective occupancies.
%The impact of correlations is stronger for these states since $E_g'$ and
%$A_{1g}$, which would be expected are partially filled.
The opposite trend is found for LiCoO$_2$ with a larger
magnitude of $\mathrm{Im}(\Sigma)$ for the $E_g$ states for the full
range of frequency shown. This suggests that for LiCoO$_2$ the
correlations have a larger impact on the nominally-unoccupied $E_g$
states since they are partially occupied via hybridization with O $p$
states, whereas the $E_g'$ and $A_{1g}$ are filled.

For LiCoO$_2$, in the high-frequency limit, $\mathrm{Re}(\Sigma)$ is
typically negative for $E_g'$ and $A_{1g}$ and positive for $E_g$.
This indicates that the static part of the correlations tend to push
$E_g'$ and $A_{1g}$ down in energy and $E_g$ up in energy as is
observed using DFT+$U$. The $U=1.9$ eV case is an exception as $J$ is
likely too large relative to $U$ in this case. The real part of the
self-energy increases at lower frequency for $E_g'$ and $A_{1g}$,
whereas it decreases for $E_g$. This leads to a higher
$\mathrm{Re}(\Sigma)$ for $E_g'$ and $A_{1g}$ than $E_g$ towards zero
frequency. Overall, the magnitude of the changes in
$\mathrm{Re}(\Sigma)$ with $U$ are significantly larger for $E_g'$ and
$A_{1g}$ than for $E_g$.

%% Is is worth commenting that the opposite trend is seen at zero
%% frequency? I am not sure we have any understanding here.

For CoO$_2$, the self-energy of the $E_g$ states has a small real part
(at most 0.21 eV), which decreases and becomes negative at low
frequency. The magnitude is substantially larger for $E_g'$ and
$A_{1g}$ than $E_g$ with a maximum magnitude of 1.2 eV for $U=5.9$ eV.
For these states, like in the LiCoO$_2$ case, the values are negative
at high frequency (except for very low $U$) and become positive at low
frequency. As opposed to the imaginary part, the real part of the
self-energy has smaller magnitude for CoO$_2$ than for LiCoO$_2$.

%This indicates that the quasiparticle levels of LiCoO$_2$
%are shifted more, but those of CoO$_2$ are broadened more.

\begin{figure}[htbp]
\begin{center}
\includegraphics[width=\linewidth]{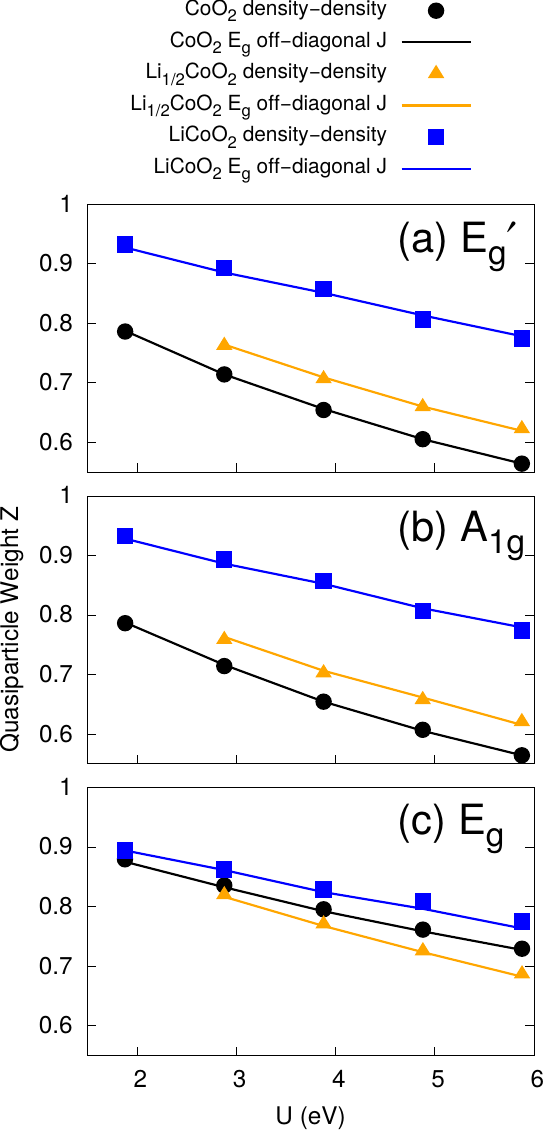}
\end{center}

\caption{Quasiparticle weight $Z$ as a function of $U$ for (a) $E_g'$,
  (b) $A_{1g}$, and (c) $E_g$ orbitals in CoO$_2$, Li$_{1/2}$CoO$_2$,
  and LiCoO$_2$.\label{fig:weights_vs_u}}
\end{figure}

From the low-frequency behavior of $\mathrm{Im}(\Sigma)$, we compute
the quasiparticle weight $Z=[1-\partial\mathrm{Im}(\Sigma)/\partial
  i\omega|_{i\omega\rightarrow0}]^{-1}$, shown in Fig.
\ref{fig:weights_vs_u}. This quantity is unity for $U=J=0$
[$\mathrm{Im}(\Sigma)=0$] and is inversely proportional to
the effective mass arising from electron interactions. All the values
decrease with $U$, as expected, in a roughly linear fashion. $Z$ is
always larger for LiCoO$_2$ than CoO$_2$, consistent with the fact
that LiCoO$_2$ is a band insulator. This effect is pronounced in the
$E_g'$ and $A_{1g}$ states, for which the CoO$_2$ values are
0.14--0.20 lower than those of LiCoO$_2$. For the $E_g$ states, the
disparity is smaller, with differences of only 0.01--0.04. For
CoO$_2$, $Z$ is larger and decreases less rapidly for the $E_g$
orbitals compared to the $E_g'$ and $A_{1g}$
orbitals. From $U=1.9$ to 5.9 eV, $Z$ of the $E_g'$ and $A_{1g}$
orbitals of CoO$_2$ goes from 0.79 to 0.57 and that of the $E_g$
orbitals goes from 0.88 to 0.73. For LiCoO$_2$, over the same range of
$U$, $Z$ of the $E_g'$ and $A_{1g}$ orbitals goes from 0.93 to 0.77
and that of the $E_g$ orbitals goes from 0.89 to 0.77. Here, $Z$ is
smaller and decreases less rapidly for the $E_g$ states such that $Z$
is the same for all the orbitals at $U=5.9$ eV.

\begin{figure}[htbp]
\begin{center}
\includegraphics[width=\linewidth]{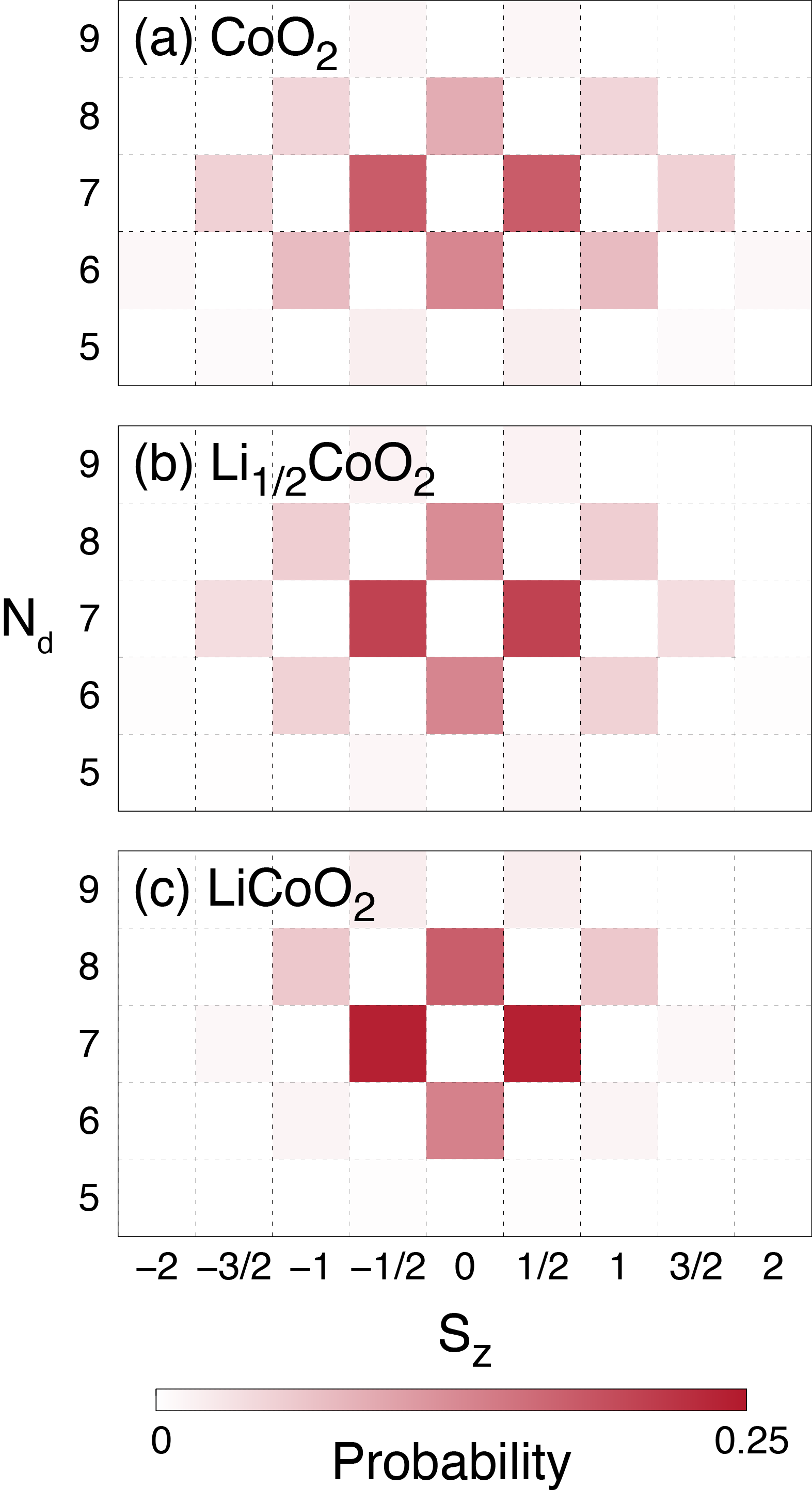}
\end{center}

\caption{Probability of Co atomic states with number of $d$ electrons
  $N_d$ and spin projection $S_z$ for (a) CoO$_2$, (b)
  Li$_{1/2}$CoO$_2$, and (c) LiCoO$_2$.
  \label{fig:probns}}
\end{figure}

\subsection{Atomic configurations and $d$ occupancies}

To further understand the detailed electronic configuration of CoO$_2$
and LiCoO$_2$, in Fig. \ref{fig:probns} we plot the probabilities of
the different atomic configurations sampled by the CTQMC solver in
terms of the number of $d$ electrons ($N_d$) and the spin projection
$S_z$. The results for $U=4.9$ eV are shown as a representative
example. We note that the probability distribution is symmetric about
$S_z=0$ since our DFT+DMFT calculations are paramagnetic (i.e., there
is no long-range magnetic order).

Although CoO$_2$ and LiCoO$_2$ are nominally $d^5$ and $d^6$,
respectively, the probability distribution is centered at higher
values of $N_d$ for both cases due to the appreciable hybridization
with O $p$ states. For example, for LiCoO$_2$ there is substantial
time in the Monte Carlo simulation in which an electron from an O $p$
state has hopped into an $E_g$ orbital, leading to a $d^7$ state.
There are substantial fluctuations in $N$ as well as $S_z$ for the Co
site in both systems. For CoO$_2$ the spin fluctuations are moderately
larger than in LiCoO$_2$; there is even probability of $S_z=3/2$
states. We note that these fluctuations of the Co site highlight why both
DFT and DFT+$U$ struggle to capture all the physics in this system.

\begin{figure*}[htbp]
\begin{center}
\includegraphics[width=\linewidth]{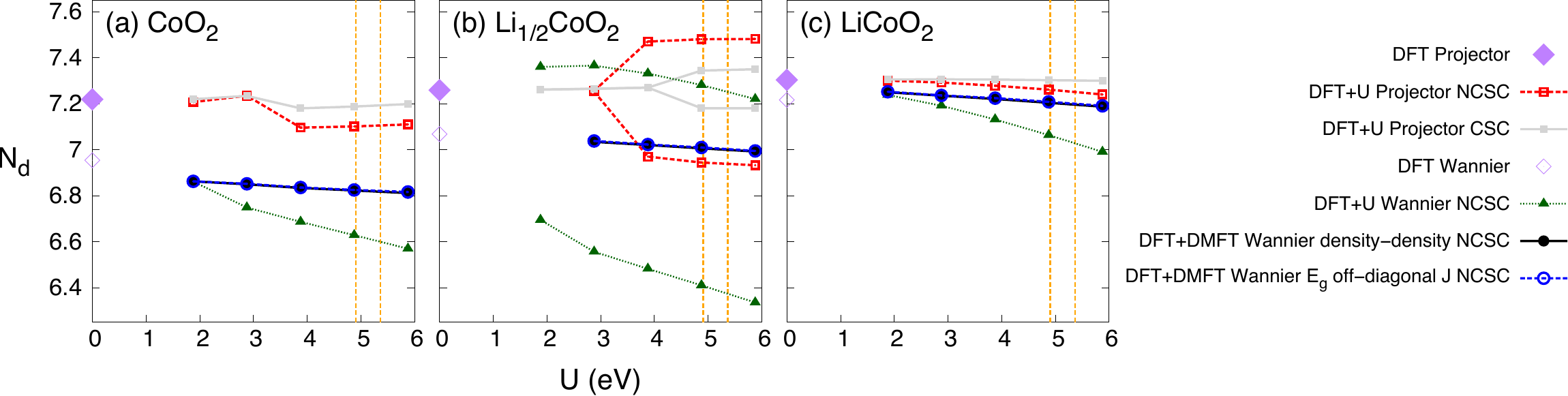}
\end{center}

\caption{$N_d$ versus $U$ for (a) CoO$_2$, (b) Li$_{1/2}$CoO$_2$, and
  (c) LiCoO$_2$ for all the methodologies employed in this study. The
  dashed orange lines indicate the computed values of $U$ for
  LiCoO$_2$ (lower value) and CoO$_2$ (higher value) within the linear
  response approach. Multiple lines in panel (b) correspond to
  different values for the two Co sites.\label{fig:nd}}
\end{figure*}

It is also useful to examine the behavior of $N_d$ versus $U$ for all
the methodologies employed in this work in Fig. \ref{fig:nd}. This has
been shown to give insight into the behavior and impact of the double
counting correction, which is purely a function of $N_d$
\cite{Wang2012195136,Dang2014125114,park_computing_2014}. Within DFT,
one can observe that $N_d$ is larger for the projector correlated
subspace than the Wannier correlated subspace. The difference is
moderate for LiCoO$_2$ (0.09), but significantly larger for CoO$_2$
(0.26). LiCoO$_2$ has 0.08 (0.26) more $d$ electrons than CoO$_2$ in
the projector (Wannier) correlated subspace. These values are much
smaller than the nominal value of unity, which is indicative of the
strong $p$--$d$ rehybridization in this system
\cite{wolverton_first-principles_1998,marianetti_role_2004}.

%% EBI: removed this for now. dc does shift Nd, but the interaction part of the functional also affects Nd
%% as dictated by the standard fully-localized limit double-counting

$N_d$ typically decreases with $U$. When treating LiCoO$_2$ with
DFT+$U$ in the Wannier correlated subspace, we find a transition to
high-spin Co at $U$ of around 4 eV. Since Co in LiCoO$_2$ is not
high-spin in experiments
\cite{van_elp_electronic_1991,menetrier_insulator-metal_1999,menetrier_really_2008},
we only consider the low-spin Co state to facilitate comparison with
DFT+DMFT, for which we do not find this spurious high-spin state. For
LiCoO$_2$ in the Wannier correlated subspace (the band insulator
state), the decrease in $N_d$ with $U$ for DFT+$U$ is around 0.25
electrons. The inclusion of dynamical correlations (DFT+DMFT)
substantially dampens this decrease to a only 0.06 electrons. Unlike
in the Wannier case, for DFT+$U$ in the projector correlated subspace,
the decrease in $N_d$ is small in magnitude (around $0.03$ electrons)
and including charge self-consistency leads to even smaller changes on
the order of 0.006 electrons.

The behavior is similar for CoO$_2$. Here, DFT+DMFT gives a small
decrease in $N_d$ of 0.04 electrons, whereas the decrease is much more
substantial (0.3 electrons) for DFT+$U$ in the Wannier correlated
subspace, which describes CoO$_2$ as an insulator except for the
smallest $U$ considered. For DFT+$U$ in the projector case, in which
there is a discontinuous decrease in $N_d$ corresponding to a
metal-insulator transition, charge self-consistency dampens the change
in $N_d$ from around 0.1 to 0.02 electrons.

\begin{figure*}[htbp]
\begin{center}
\includegraphics[width=\linewidth]{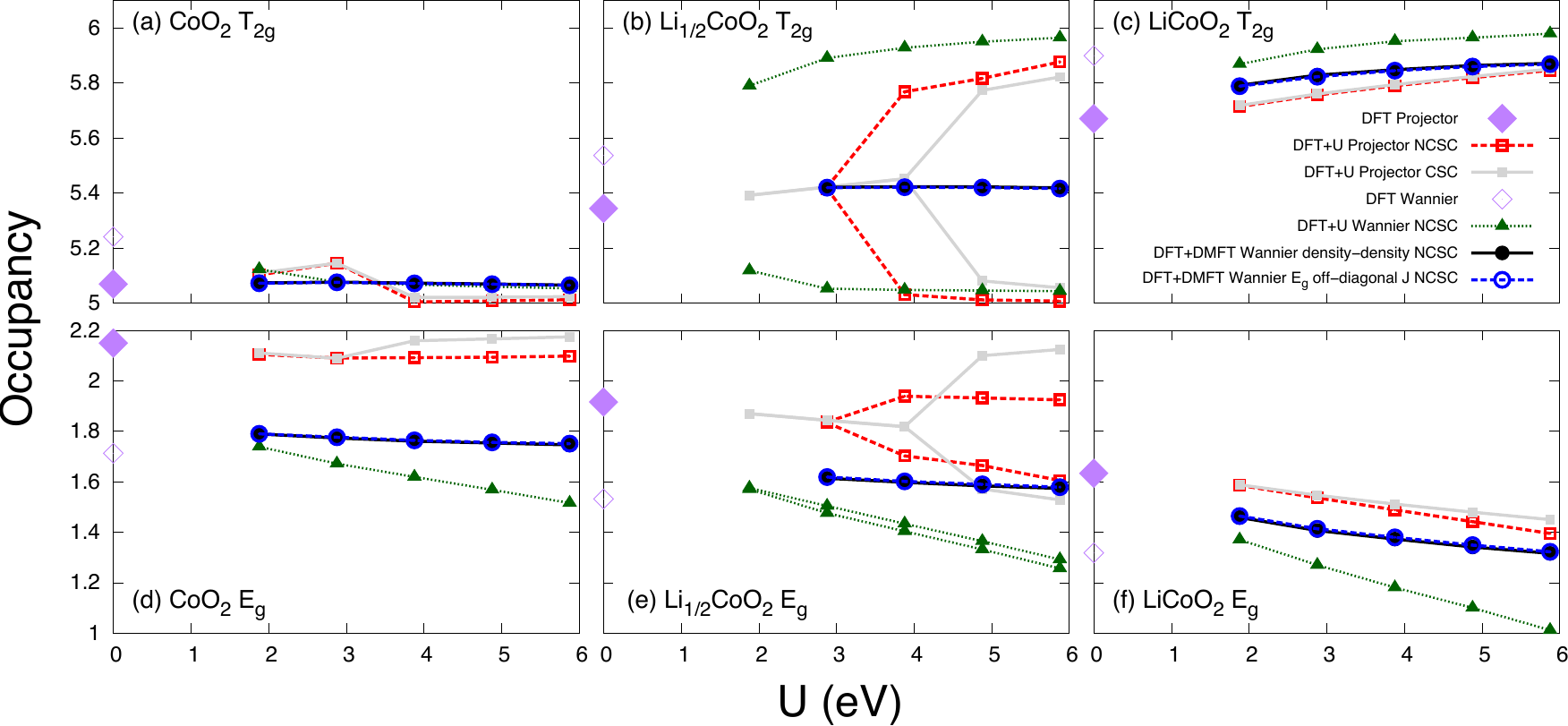}
\end{center}

\caption{Total $T_{2g}$ occupancy versus $U$ for (a) CoO$_2$, (b)
  Li$_{1/2}$CoO$_2$, and (c) LiCoO$_2$ for all the methodologies
  employed in this study. (d), (e), and (f) show the corresponding
  $E_g$ plots. Multiple lines in panels (b) and (e) correspond to
  different values for the two Co sites.\label{fig:t2g_eg_vs_u}}
\end{figure*}

The behavior of $N_d$ versus $U$ can further be understood by
decomposing $N_d$ into the components from the $T_{2g}$ ($E_g'$ and
$A_{1g}$) and $E_g$ orbitals, as shown in Fig. \ref{fig:t2g_eg_vs_u}.
We do not show the individual $E_g'$ and $A_{1g}$ occupancies for
brevity, but they are included in the Supplemental
Material \footnote{See Supplemental Material for crystal structure
  details and a breakdown of $N_d$ into the $E_g'$, $A_{1g}$, and
  $E_g$ occupancies.}. Within DFT, the Wannier correlated subspace
leads to higher (lower) occupancy of $T_{2g}$ ($E_g$) by 0.13--0.23
(0.32--0.39) electrons compared to the projector case. In DFT+$U$, the
LiCoO$_2$ $T_{2g}$ occupancy increases with $U$, whereas the $E_g$
occupancy decreases more rapidly; this leads to the overall decrease
in $N_d$. For the Wannier case, the $T_{2g}$ occupancy increases more
rapidly with $U$ at lower $U$ compared to the projector case; for
larger $U$, the occupancy begins to saturate close to the nominal
value of 6. Similarly, the decrease in $E_g$ occupancy is more
substantial in the Wannier case compared to the projector case.
Including charge self-consistency has a very small effect on the
occupancies of LiCoO$_2$ in the projector case. The trends in the
occupancies are the same for DFT+DMFT as in DFT+$U$, but the magnitude
of the changes in occupancy are much smaller.

Within DFT+$U$, the CoO$_2$ $T_{2g}$ occupancy is relatively constant
with $U$ in the Wannier correlated subspace, whereas the $E_g$
occupancy decreases by 0.22 electrons. The CoO$_2$ $T_{2g}$ occupancy
also does not change very much with $U$ in the projector case, though
there is a discontinuity leading to a decrease in occupancy through
the metal-insulator transition. In this case, charge self-consistency
serves to slightly enhance the $T_{2g}$ and $E_g$ occupancies. As in
the case of LiCoO$_2$, the changes in occupancy within DFT+DMFT are
smaller than those of DFT+$U$ with both $T_{2g}$ and $E_g$ occupancies
slightly decreasing by 0.003 and 0.04, respectively, over the range of
$U$. We note that the DFT+DMFT occupancies are fairly similar to those
of DFT for both CoO$_2$ and LiCoO$_2$.

%% Dynamical correlations appear to dampen the changes in $N_d$ in the
%% same fashion as charge self-consistency.

%% Overall, the changes are more moderate than those of
%% the Wannier case, in which the $E_g$ occupancy is nearly constant with
%% a range of only 0.02 electrons.
\subsection{Electronic structure of Li$_{1/2}$CoO$_2$}
\label{sec:es_x12}
Here, we discuss the electronic structure of Li$_{1/2}$CoO$_2$, which
warrants extra attention due to the issue of spurious charge ordering
which occurs in DFT+$U$ but is not seen experimentally
\cite{takahashi_single-crystal_2007}. This known structure has an
in-plane ordering of Li corresponding to a primitive unit cell with
two formula units
\cite{reimers_electrochemical_1992,van_der_ven_first-principles_1998,wolverton_first-principles_1998}.
We perform DFT+DMFT calculations in two different ways. First, we
enforce the symmetry between the two structurally-equivalent Co sites,
i.e., only a single impurity problem is solved. However, the
aforementioned approach does not allow for charge ordering to
spontaneously break point symmetry. Therefore, we also use a second
approach where two impurity calculations are employed (i.e., one for
each Co atom in the unit cell), and we only execute this at $U=3.9$ eV
due to computational expense. Using the second approach, we do find a
stable charge-ordered state, with a difference in $N_d$ on the two
sites of 0.4 electrons, but it is slightly higher in energy than the
non-charge-ordered state (by 3 meV/f.u.). Although this is a quite
small energy difference, it it clear that DFT+DMFT removes the strong
tendency for spurious charge ordering that is produced by DFT+$U$, and
we proceed with our analysis of the symmetric solution.

Within DFT, Li$_{1/2}$CoO$_2$ is metallic with half a hole in the
$T_{2g}$ manifold, as can be seen from the density of states in Fig.
\ref{fig:dos_structure}(b). For DFT+$U$ in the Wannier correlated
subspace, like in the LiCoO$_2$ case (discussed above), we ignore
states containing high-spin Co found at large $U$; instead, we
consider the low-spin ground state, a charge-ordered insulator (see
Figs. \ref{fig:nd}(b), \ref{fig:t2g_eg_vs_u}(b), and
\ref{fig:t2g_eg_vs_u}(e)). Within DFT+$U$ in the projector correlated
subspace, Li$_{1/2}$CoO$_2$ is in a symmetric low-spin state at low
$U$ that transitions to a low-spin charge ordered insulator at larger
$U$. The DFT+DMFT self-energy in Fig. \ref{fig:sigma_densdens}
illustrates that Li$_{1/2}$CoO$_2$ is, like CoO$_2$, a Fermi liquid
with correlations most significantly affecting the $T_{2g}$ manifold.
We find a Fermi liquid up to the highest $U$ value considered, in
agreement with the metallic behavior observed in experiment
\cite{motohashi_electronic_2009,miyoshi_magnetic_2010,ou-yang_electronic_2012}.
The $T_{2g}$ quasiparticle weights, in Fig. \ref{fig:weights_vs_u} are
significantly lower than those of LiCoO$_2$ and slightly larger than
those of CoO$_2$. This reflects that the electronic structure of
Li$_{1/2}$CoO$_2$ is closer to that of CoO$_2$ than LiCoO$_2$.

As for the endmembers ($x=0$ and $x=1$), for Li$_{1/2}$CoO$_2$, $N_d$
(Fig. \ref{fig:nd}) is larger in the projector correlated subspace.
Within DFT+DMFT, $N_d$ decreases very mildly with $U$ similar to the
endmember behavior. Here, dynamical correlations only slightly reduce
$N_d$ (on the order of 0.02 electrons). Within NCSC DFT+$U$, the
magnitude of the charge ordering is substantial (as much as $\sim 0.8$
electrons) in the Wannier correlated subspace. It is significantly
smaller for the projector case, and is further dampened by charge
self-consistency. Similar effects are found for the individual
$T_{2g}$ and $E_g$ occupancies are shown in Fig.
\ref{fig:t2g_eg_vs_u}.

In summary, DFT+DMFT can properly describe the electronic structure of
Li$_{1/2}$CoO$_2$. Unlike DFT+$U$, which predicts a charge ordered
insulator, DFT+DMFT properly describes the electronic structure as a
Fermi liquid without a strong tendency for charge ordering.

\subsection{Total energy of LiCoO$_2$ and CoO$_2$}

\begin{figure}[htbp]
\begin{center}
\includegraphics[width=\linewidth]{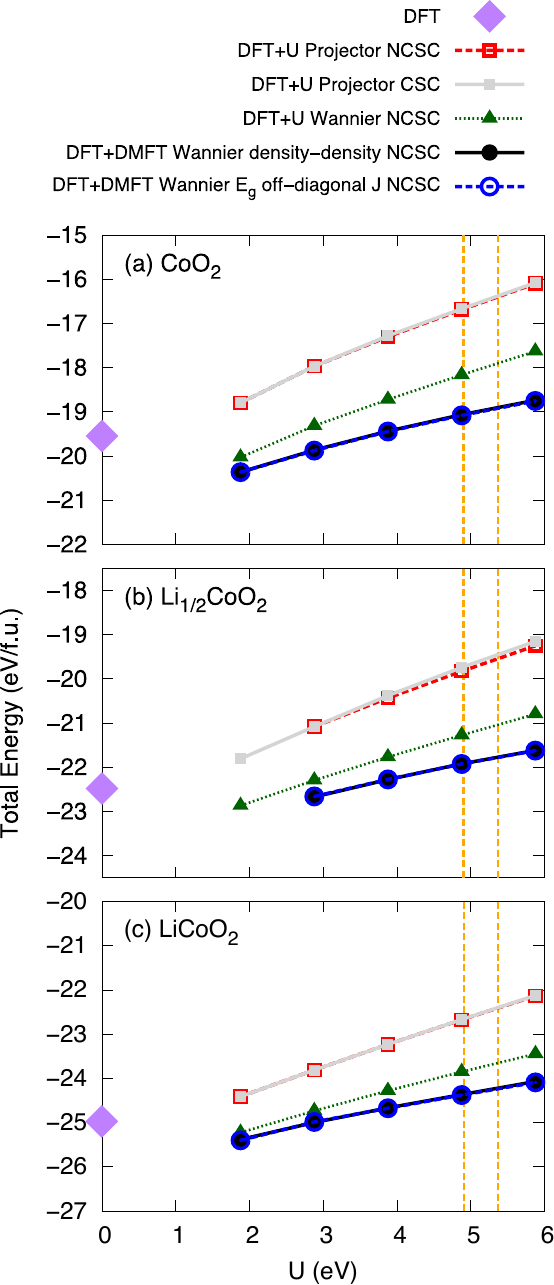}
\end{center}
\caption{Total energy of (a) CoO$_2$, (b) Li$_{1/2}$CoO$_2$, and (c)
  LiCoO$_2$, respectively, as a function of $U$ for several
  methodologies including DFT, DFT+$U$, and DFT+DMFT ($J$ remains unchanged). The dashed
  orange lines indicate the computed values of $U$ for LiCoO$_2$
  (lower value) and CoO$_2$ (higher value) within the linear response
  approach.\label{fig:energies}}
\end{figure}

Having documented the basic electronic structure, we proceed to
explore the total energy of LiCoO$_2$ and CoO$_2$ as a function of
$U$, allowing for a direct comparison between DFT+DMFT and DFT+$U$.
The total energy of LCO is shown as a function of $U$ for several
methodologies in Fig. \ref{fig:energies}. The two vertical dashed
lines indicate the values of $U$ for CoO$_2$ and LiCoO$_2$ as computed
from first principles via linear response
\cite{zhou_first-principles_2004}. The total energies increase with
$U$, as expected.
We note that extrapolating any of the results to $U=0$ does not approach
the DFT result (large purple diamond), which is simply due to the fact that
we used a fixed value of $J$ for all calculations.
%
%We note that for $U=0$ the total energies within
%DFT+$U$ and DFT+DMFT are not equal to those of DFT since we have
%chosen a fixed finite $J$; the results for larger $U$ should be
%considered the most physical.
The magnitude of the increase in total
energy with $U$ is generally greater for CoO$_2$ than LiCoO$_2$, which
makes sense since the impact of the on-site interaction is expected to
be larger for the system for which $T_{2g}$ is partially filled
(nominally). For NCSC DFT+$U$ in the Wannier correlated subspace, for
example, over the full range of $U$ shown the increase in energy of
CoO$_2$ is 2.4 eV as compared to only 1.8 eV for LiCoO$_2$. For the
same set of calculations using the projector correlated subspace, we
find the same trend with energy increases of roughly 2.7 eV for
CoO$_2$ and 2.3 eV for LiCoO$_2$. We note that the individual total
energies from methods utilizing these different correlated subspaces
(projector and Wannier) are not directly comparable, but the behavior
of the total energy with $U$ is similar.

%% LiCoO$_2$ is described as a band insulator within all of our DFT+$U$
%% results for the range of $U$ shown. In the Wannier correlated
%% subspace, we find CoO$_2$ is a non-spin-polarized metal; only for $U$
%% above 6.9 eV does CoO$_2$ transition to a magnetic insulator.
%% Similarly, we find CoO$_2$ is a non-spin-polarized metal in the
%% projector correlated subspace.

%% In
%% contrast, in the projector correlated subspace, CoO$_2$ is a
%% spin-polarized metal for smaller values of $U$. This is possibly an
%% artifact of a too-large value of $J$ in this regime of $U$. Above
%% $U=2.9$ eV, CoO$_2$ is a non-spin-polarized metal, consistent with the
%% Wannier result.

We estimate the magnitude of the error associated with neglecting
charge self-consistency via DFT+$U$ calculations in the projector
correlated subspace. Here, for LiCoO$_2$, we find only very small
differences (at most 9 meV/f.u.) between the NCSC and CSC total
energies. For CoO$_2$, the situation is similar with differences in
total energy of at most 20 meV/f.u.. The small impact of changes in
charge density on the total energies suggest the fixed charge density
should be a reasonable approximation for DFT+DMFT. More importantly,
we have a clear guideline on the magnitude of the effect for charge
self-consistency within DFT+$U$, and we expect this is an upper bound
for DFT+DMFT calculations in which the impact will likely be dampened.

%% However, for $U>2.9$ eV, CoO$_2$ orbitally orders and
%% opens up a band gap; in this regime the total energies are lowered by
%% several hundreds of meV/f.u. compared to those of the NCSC
%% calculations.

%While
%charge self-consistency does significantly affect the energetics of
%CoO$_2$ upon opening a band gap, this long-range-ordered insulating
%state is likely an artifact of the crude Hartree-Fock treatment.

Within DFT+DMFT, we find very little impact of including the
off-diagonal $J$ interaction terms within the $E_g$ manifold in
addition to the density-density interactions. The magnitude of the
differences is typically only around 5--15 meV/f.u. for LiCoO$_2$ and
3--9 meV/f.u. for CoO$_2$. This suggests that density-density
interactions are likely sufficient to describe this class of systems.
In all of the following results, we find no significant difference in
employing these two interaction forms. The DFT+DMFT results, which
employ the Wannier correlated subspace, appear to merge with the
corresponding DFT+$U$ results in the limit of small $U$, as should be
the case. We note again that in this limit, neither the DFT+$U$ nor
the DFT+DMFT results recover the DFT values (large purple diamonds
in Fig. \ref{fig:energies}) simply because we take have taken a fixed
finite $J$ value, whereas the DFT values correspond to $J=0$.

%% 2.4 1.8 Wannier

We find the general impact of dynamical correlations on the energetics
is to dampen the magnitude of the increase in total energy with $U$ as
compared to the static Hartree-Fock treatment in DFT+$U$. When $U$ is
increased from 1.9 to 5.9 eV, the total energy of LiCoO$_2$ increases
by 1.8 eV within DFT+$U$ as opposed to only 1.3 eV within DFT+DMFT.
For CoO$_2$, the magnitude of these energies is substantially larger
with an increase of 2.4 eV for DFT+$U$ and 1.6 eV for DFT+DMFT. By
this measure, dynamical correlations decrease the energy penalty of
$U$ by 26\% for LiCoO$_2$ and 33\% for CoO$_2$. Therefore, dynamical
correlations have a larger impact on CoO$_2$ than LiCoO$_2$. This
corresponds to very large absolute differences in the energies
predicted by DFT+$U$ and DFT+DMFT. For CoO$_2$, for example, around
the linear response values of $U$ the difference in energy is around 1
eV. This strongly suggests dynamical correlations, missing in the
DFT+$U$ approach, are important for accurate total energies.

It should be emphasized that the difference between DFT+DMFT and
DFT+$U$ changes substantially as a function of $x$, and this error
will therefore strongly affect observables. One strategy to correct
errors within DFT+$U$ calculations is to tune $U$ to artificially low
values, and this can be successful if $U$ is first calibrated to some
experimental observable. However, our work indicates that the errors
vary strongly with composition, and therefore DFT+$U$ studies of
compositional phase stability would need to tune $U$ as a function of
doping, a far more challenging task. Below we explore the average
battery voltage, where the composition dependent errors within DFT+$U$
have severe consequences.

\subsection{Average intercalation voltage}

\begin{figure}[htbp]
\begin{center}
\includegraphics[width=\linewidth]{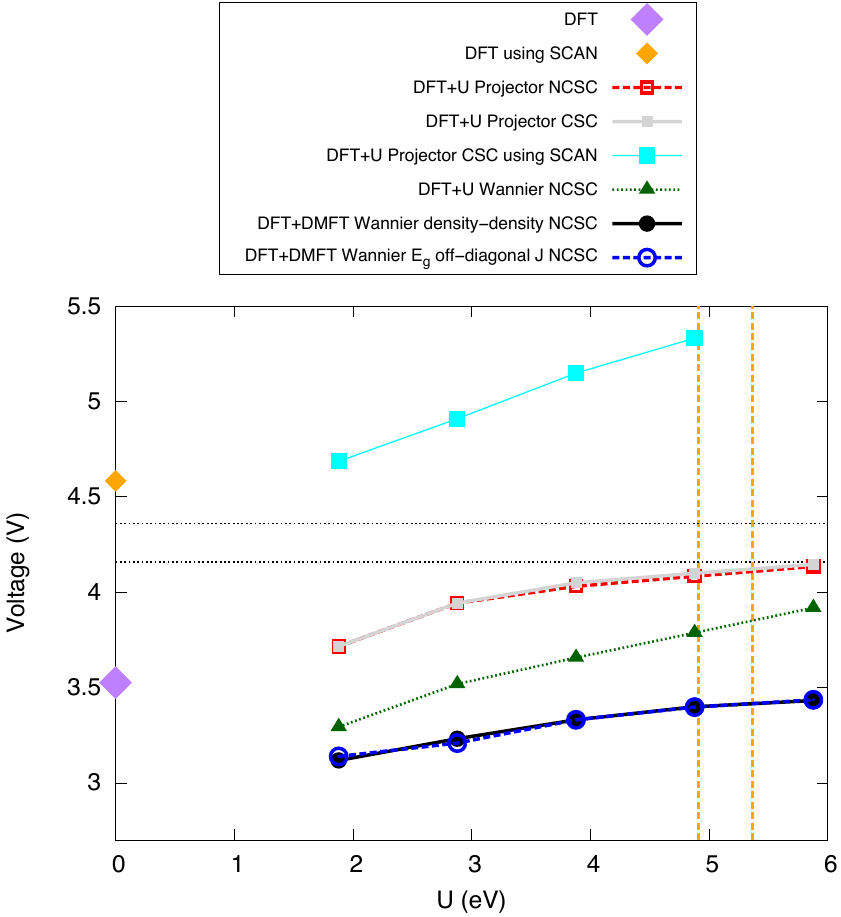}
\end{center}
\caption{Computed intercalation voltage of LCO via DFT, DFT+$U$, and
  DFT+DMFT as a function of $U$. The dashed orange lines indicate the
  computed values of $U$ for LiCoO$_2$ (lower value) and CoO$_2$
  (higher value) within the linear response approach. The dotted black
  lines indicate the expected range of the experimental result
  \cite{amatucci_coo2_1996}.\label{fig:intercalation_voltage}}
\end{figure}

We turn our attention to the average intercalation voltage of LCO for
$0\le x\le 1$, plotted in Fig. \ref{fig:intercalation_voltage}, which
is a key observable for a rechargeable battery cathode. The average
intercalation voltage $V$ is computed via
$eV=E(\mathrm{Li})+E(\mathrm{CoO_2})-E(\mathrm{LiCoO_2}),$ where $e$
is the elementary charge and body-centered-cubic Li is the reference
electrode \cite{aydinol_ab_1997}. As has been known, DFT tends to
underpredict the experimental voltage \cite{aydinol_ab_1997}, in this
case by around 0.7 V. The predicted voltage increases with DFT+$U$. In
the projector correlated subspace, the value comes close to reasonable
agreement with the experimental value (approximately 4.26 V
\cite{amatucci_coo2_1996}) for the largest $U$ considered, and charge
self-consistency has a negligible effect. The voltage also becomes
larger than the DFT value for DFT+$U$ in the Wannier case, though the
predicted values are appreciably smaller and do not reach agreement
with experiment. In stark contrast, within DFT+DMFT, the predicted
intercalation voltage is smaller than the DFT value up to the highest
computed value of $U=6$ eV. As compared to DFT+$U$, the voltage at
fixed $J$ increases much more slowly as a function of $U$. This mainly
stems from the dampened increase in energy for CoO$_2$. For the
computed values of $U$, the predicted voltage is only 3.39--3.45 V, a
moderate amount less than that of pure DFT.

The increase in $V$ within DFT+$U$ and general agreement with
experiment was shown previously \cite{zhou_first-principles_2004} and
seemed to suggest that DFT+$U$ is reliable for this class of
materials. However, this viewpoint should be carefully scrutinized
given that DFT+$U$ is a rather crude theory, in that DFT+$U$ is
obtained from DFT+DMFT when the quantum impurity problem is solved
within static mean-field theory, neglecting dynamical correlations.
Therefore, DFT+DMFT is superior in every respect. Since the voltage
curve predicted by DFT+DMFT produces a result smaller than that of
DFT, opposite to that of DFT+$U$, dynamical correlations are clearly
essential to describe the energetics of LCO. The fact that DFT+$U$
increases the voltage, relative to DFT, and provides more reasonable
agreement with experiment appears to be fortuitous.

Given that DFT+DMFT actually worsens the predicted voltage as compared
to DFT, we are left with the puzzling question as to why. We explore
several possibilities. First and foremost, while DMFT should improve
DFT with respect to local physics, it is possible that there are still
substantial nonlocal errors within the density functional employed in
the DFT+DMFT functional. Given the recent successes of the relatively
new SCAN functional
\cite{isaacs_performance_2018,zhang_efficient_2018}, which contains
nonlocal physics via a dependence on the orbital kinetic energy
density, an obvious question is how the voltage would change if we
replaced the PBE functional with SCAN; we pursue this idea at the
level of DFT+$U$ (see Fig. \ref{fig:intercalation_voltage}). The $U=0$
voltage predicted by SCAN is 4.58 V, which is already \emph{greater}
than the experimental voltage, in stark contrast to LDA and PBE. One
expects the trends as a function of $U$ for DFT(SCAN)+$U$ and
DFT(SCAN)+DMFT to be unchanged relative to PBE given that the SCAN
electronic structure is very similar to that of PBE. As expected,
increasing $U$ within DFT(SCAN)+$U$ causes the voltage to further
increase, moving away from the experimental value. We do not compute
the DFT(SCAN)+DMFT results due to the computational cost, but we
anticipate that they should have the same $U$-dependence as
DFT(PBE)+DMFT, just as DFT(SCAN)+$U$ and DFT(PBE)+$U$ have a very
similar $U$-dependence. If so, the DFT(SCAN)+DMFT voltage should be
mildly decreased compared to the DFT(SCAN) voltage, yielding
reasonable agreement with experiment. Alternatively, DFT(SCAN)+$U$
only worsens the voltage prediction, and this suggests the reason why
DFT(PBE)+$U$ performs well is due to the cancellations of two large
and distinct errors. Therefore, our results suggest one cannot expect
DFT+$U$ to perform as a predictive tool in the context of
compositional phase stability.

Of course, another factor to consider is that the precise value of the predicted voltage will depend on
the precise values $U$ and $J$. Allowing for small $U$
differences in the endmembers will result in small changes in the voltage.
Similiar arguments might be made for other uncertainties in the methodology, such as the double
counting correction, etc. We do not attempt to build a case for the
most correct set of parameters in this work. It is worth noting that
one early explanation for discrepancies in the DFT predicted voltages
was the overestimated magnitude of the cohesive energy of
body-centered-cubic Li within LDA \cite{aydinol_ab_1997}. However, we
find the cohesive energy predicted by PBE ($-1.60$ eV) and SCAN
($-1.59$ eV) are in good agreement with experiment ($-1.63$ eV
\cite{kittel_introduction_1986}). Therefore, the discrepancy between
theory and experiment should stem from the energetics of the cathode
material itself.

\subsection{Phase stability of Li$_{1/2}$CoO$_2$}
\begin{figure}[htbp]
\begin{center}
\includegraphics[width=\linewidth]{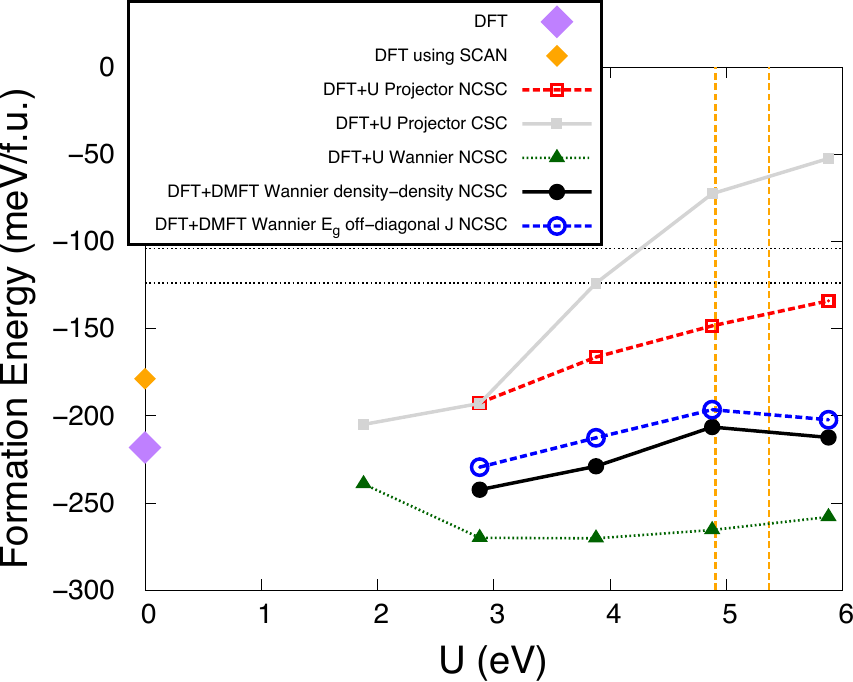}
\end{center}
\caption{Computed formation energy of Li$_{1/2}$CoO$_2$ via DFT,
  DFT+$U$, and DFT+DMFT as a function of $U$. The dashed orange lines
  indicate the computed values of $U$ for LiCoO$_2$ (lower value) and
  CoO$_2$ (higher value) within the linear response approach. The
  dotted black lines indicate the expected range of the experimental
  result \cite{amatucci_coo2_1996}.\label{fig:fe}}
\end{figure}

As another test of the computed DFT+DMFT total energy, we compute the
phase stability of Li$_{1/2}$CoO$_2$. The formation energy $\Delta E$,
computed as $E(\mathrm{Li_{1/2}CoO_2}) - \frac{1}{2}[E(\mathrm{CoO_2})
  + E(\mathrm{LiCoO_2})]$, is shown in Fig. \ref{fig:fe}. We use the
difference in average experimental voltage values for $0<x<1/2$
($V_-$) and $1/2<x<1$ ($V_+$) to estimate the experimental formation
energy for $x=1/2$, via $\Delta E = x(1-x)(eV_+-eV_-)$
\cite{aykol_local_2014}. Using the data of Ref.
\citenum{amatucci_coo2_1996}, we compute $\Delta E$ of $-114$ meV/f.u.
for Li$_{1/2}$CoO$_2$. As the formation energy has a significantly
smaller scale than the voltage, it is more sensitive to the different
methodologies employed in this work. Furthermore, the energetics of
NCSC calculations will be less reliable when there is a substantial rearrangement of charge.

We find a DFT value of $-218$ meV/f.u. NCSC DFT+$U$ in the projector
correlated subspace significantly decreases the magnitude of the
formation energy and nearly reaches agreement with experiment at large
$U$. Charge self-consistency tends to destabilize Li$_{1/2}$CoO$_2$
compared to the endmembers, further lowering the formation energy
magnitude to values much smaller than experiment. While the formation
energy remains negative (in agreement with the experimentally-known
phase stability), we have shown previously that the more common
DFT+$U$ methodology based on spin-dependent DFT incorrectly predicts
phase separation of Li$_{1/2}$CoO$_2$ in the absence of charge
ordering \cite{isaacs_compositional_2017}. The formation energy is
also significantly affected by $U$ using DFT+$U$ in the Wannier
correlated subspace, but in the opposite direction. In this case, the
formation energy is pushed more negative to values of 260--270
meV/f.u.

Given that DFT+DMFT prefers not to charge order,
the errors associated with NCSC are expected to be comparable to those estimated
for the endmembers, which would still be small on this scale.
%In stark contrast to DFT+$U$,
DFT+DMFT shows only a small increase in
formation energy with $U$, on the order of tens of meV/f.u. The
predicted value in the range of computed $U$ is similar to that of
DFT.
%This results from the fact that dynamical correlations
%significantly lower the energy of Li$_{1/2}$CoO$_2$, as shown in Fig.
%\ref{fig:energies}, by an amount greater than the average of those of
%$x=0$ and $x=1$. This represents further evidence of the importance of
%dynamical correlations to describe the thermodynamics of LCO.
%Comparing to the corresponding DFT+$U$ calculations in the Wannier
%correlated subspace suggests such correlations in this case serve to
%enhance the phase stability of $x=1/2$.
The DFT+DMFT error may also
 be associated in part with the DFT exchange-correlation
functional. We also plot the formation energy using the SCAN functional,
which is approximately 50 meV/f.u. higher than the PBE value. Therefore, we
would anticipate the DFT(SCAN)+DMFT results to be shifted by roughly
the same amount, which would be much closer to the experimental range. In summary, similar to the voltage behavior, our results suggest that dynamical correlations significantly impact the predicted formation energy and that DFT+DMFT based on SCAN will lead to improved agreement with experiment.

\section{Conclusions}

We investigate the electronic structure, intercalation voltage, and
phase stability of LCO with the many-body DFT+DMFT methodology and
compare to DFT and DFT+$U$. In DFT+DMFT, LiCoO$_2$ is a band
insulator, while we find that CoO$_2$ and Li$_{1/2}$CoO$_2$ are
moderately correlated Fermi liquids, without a strong tendency for
charge ordering in Li$_{1/2}$CoO$_2$, in agreement with experiments.
Dynamical correlations (missing in DFT+$U$) substantially impact the
energetics of LCO by dampening the changes in total energy and $N_d$
found via the DFT+$U$ approach, especially for CoO$_2$ and
Li$_{1/2}$CoO$_2$. The intercalation voltage behavior of DFT+$U$ and
DFT+DMFT are qualitatively different, with the latter decreasing the
voltage with respect to DFT; the phase stability of Li$_{1/2}$CoO$_2$
within DFT+DMFT also differs starkly from DFT+$U$. DFT+DMFT
calculations based on the PBE GGA DFT functional underpredict the
voltage and overestimate the stability of Li$_{1/2}$CoO$_2$ compared
to experiment; we find evidence that DFT+DMFT based on the more
accurate SCAN DFT functional will lead to significantly closer
agreement to experiment.

We find that dynamical correlations are important to describe this
class of materials despite the relatively modest quasiparticle
weights. Our results suggest that the Hartree-Fock treatment of the
impurity problem in DFT+$U$ is insufficient to accurately describe the
electronic structure and thermodynamics of correlated
electron materials. In addition, due to the strong composition
dependence of the impact of dynamical correlations, our results
challenge the common practice of artificially tuning $U$ within
DFT+$U$ to compensate for the errors of Hartree-Fock. Given the
significant computational expense of solving the impurity problem in
DFT+DMFT, the development of less computationally expensive but still
sufficiently accurate impurity solvers will be important future work
to enable the study of compositional phase stability of strongly
correlated electron materials.

%% (captured by DFT+DMFT, but missing
%% in DFT+$U$)

\section{Acknowledgments}
We acknowledge Hyowon Park (UIC/Argonne) for useful discussions. This
research used resources of the National Energy Research Scientific
Computing Center, a DOE Office of Science User Facility supported by
the Office of Science of the U.S. Department of Energy under Contract
No. DE-AC02-05CH11231. E.B.I. gratefully acknowledges the U.S.
Department of Energy Computational Science Graduate Fellowship (Grant
No. DE-FG02-97ER25308) for support. C.A.M. was supported by the grant
DE-SC0016507 funded by the U.S. Department of Energy, Office of
Science.

\bibliography{main}

\end{document}